\documentclass[aps,prl,twocolumn,amsmath,amssymb,showpacs,showkeys]{revtex4-1}
\usepackage{graphicx}
\begin{document}
\title{Magnetic anisotropy of thin sputtered MgB$_{2}$ films on MgO substrates in high magnetic fields} 
\author{Savio Fabretti}
\email[]{fabretti@physik.uni-bielefeld.de}
\author{Inga-Mareen Imort}
\author{Timo Kuschel}
\affiliation{Thin films and physics of nanostructures, Bielefeld University, Germany}
\author{Thomas Dahm}
\affiliation{Condensed matter theory group, Bielefeld University, Germany}
\author{Veerendra K. Guduru}
\author{Uli Zeitler}
\affiliation{High Field Magnet Laboratory and Institute of Molecules and Materials, Radboud University Nijmegen, The Netherlands.}

\author{Andy Thomas}
\homepage[]{www.spinelectronics.de}
\affiliation{Thin films and physics of nanostructures, Bielefeld University, Germany and Institut f\"ur Physik, Johannes Gutenberg Universit\"at Mainz, Germany}
\date{\today}
\begin{abstract}
We investigated the magnetic anisotropy ratio of thin sputtered polycrystalline MgB$_{2}$ films on MgO substrates. Using high magnetic field measurements, we estimated an anisotropy ratio of 1.35 for T=0\,K with an upper critical field of 31.74\,T in the parallel case and 23.5\,T in the perpendicular case. Direct measurements of a magnetic-field sweep at 4.2\,K show a linear behavior, confirmed by a linear fit for magnetic fields perpendicular to the film plane. Furthermore, we observed a change of up to 12\% of the anisotropy ratio in dependence of the film thickness.
\end{abstract}
\maketitle
%
%
%
One possible application for superconductors is the use as a spin detector in ferromagnet/ insulator/ superconductor tunneling junctions. The so-called Meservey-Tedrow method (M-T method) offers the possibility to measure the spin polarization of the tunneling electrons due to the different Zeeman energies of the spin up and spin down electrons \cite{Meservey:1994ug,Tedrow:1971tj}. In particular, the superconductor requires a moderate transition temperature, a large penetration depth, and a composition of light elements. 

The moderate transition temperature leads to an energy gap larger than the Zeeman energy in magnetic fields of approximately 3\,T. The large penetration depth allows the magnetic field to enter the superconductor homogeneously, while the magnetic field is aligned in parallel to magnetization of the ferromagnet, i.e. in the film plane. The spin orbit scattering of the quasi particles scales with the fourth power of the atomic number. Until now, aluminum has been used as the superconducting spin detector, because it fulfills all the described requirements. In particular, there are few compounds consisting of only lighter elements (compared to Al). 

Consequently, Magnesiumdiboride (MgB$_2$) is an attractive candidate for M-T tunneling experiments, because both constituents are lighter than aluminum. Furthermore, its larger penetration depth ($\lambda_{\textrm{Al}}=15.7\,\textrm{nm}, \lambda_{\textrm{MgB2}}>30\,\textrm{nm}$ \cite{Hauser:1972kj,Moshchalkov:2009th}) would allow a measurement of perpendicularly magnetized materials \cite{Balke:2007eb,Kugler:2011dz}, if the junctions structures are defined well below 50\,nm. On top of that, MgB$_{2}$ has been a topic of interest for technical applications ever since the discovery of its superconductivity more than 10 years ago \cite{Nagamatsu:2001bv,Vinod:2007wm}. 

In pursuit of a suitable measurement method for these materials, we investigated the magnetic anisotropy and tunneling properties of thin MgB$_{2}$ films sputtered onto cubic (001) MgO substrates. Only sputtering is suitable for the large-scale production of spintronic devices, where MgO is the most used substrate for, e.g., half metallic Heusler compounds \cite{Graf:2011jj,Ebke:2010ci}. Unfortunately, sputtered MgB$_{2}$ films are polycrystalline because of the high vapor pressure of magnesium during the sputtering process \cite{Liu:2001uga}. This characteristic requires the measurement of the anisotropy ratio of the MgB$_{2}$ as an indication of the texture of the films.

The magnetic anisotropy ratio $\gamma$ of thin MgB$_{2}$ films varies between 1.25 and 2.0 \cite{Buzea:2001vt}. In 2001, Jung et al. investigated the magnetic anisotropy of thin epitaxial MgB$_{2}$ films and determined the upper magnetic field to be 24$\pm$3\,T for the c-plane direction and 30$\pm$2\,T for the a-b-plane direction \cite{Jung:2001wf}. They estimated the upper magnetic field using field-cooling measurements in field strengths up to 20\,T and pulsed fields of up to 60\,T. With these measurements, they were able to show that their epitaxial films, which were fabricated using pulsed laser deposition, were in the clean limit. Investigations of samples in the dirty limit from Patnaik and Gurevich showed a strong increase of the upper critical field dependent on impurity scattering \cite{Gurevich:2003dd, Patnaik:2001vd}.

In this work, we will present the direct measurement of the upper critical fields in field-cooling experiments and compare our results of a high-field magnetic sweep measurement (0-33\,T) at a constant temperature of 4.2\,K. Furthermore, we will show that for polycrystalline films, a linear approximation for the anisotropy ratio down to 4.2\,K is in good agreement with the experimental data. Up to now, only high field measurements of MgB2 films not suitable for M-T investigations ($>$400\,nm) are available \cite{Gurevich:2003dd}.

%
%
We fabricated MgB$_{2}$ films with thicknesses ranging from 30\,nm to 120\,nm using magnetron RF-DC co-sputtering on MgO substrates. During the co-sputtering process, each sample was heated to a substrate temperature of T=$288^{\circ}$C and subsequently subjected to an in-situ annealing process at  $600^{\circ}$C for 30\,min. 
\begin{figure}
\includegraphics[width=8cm]{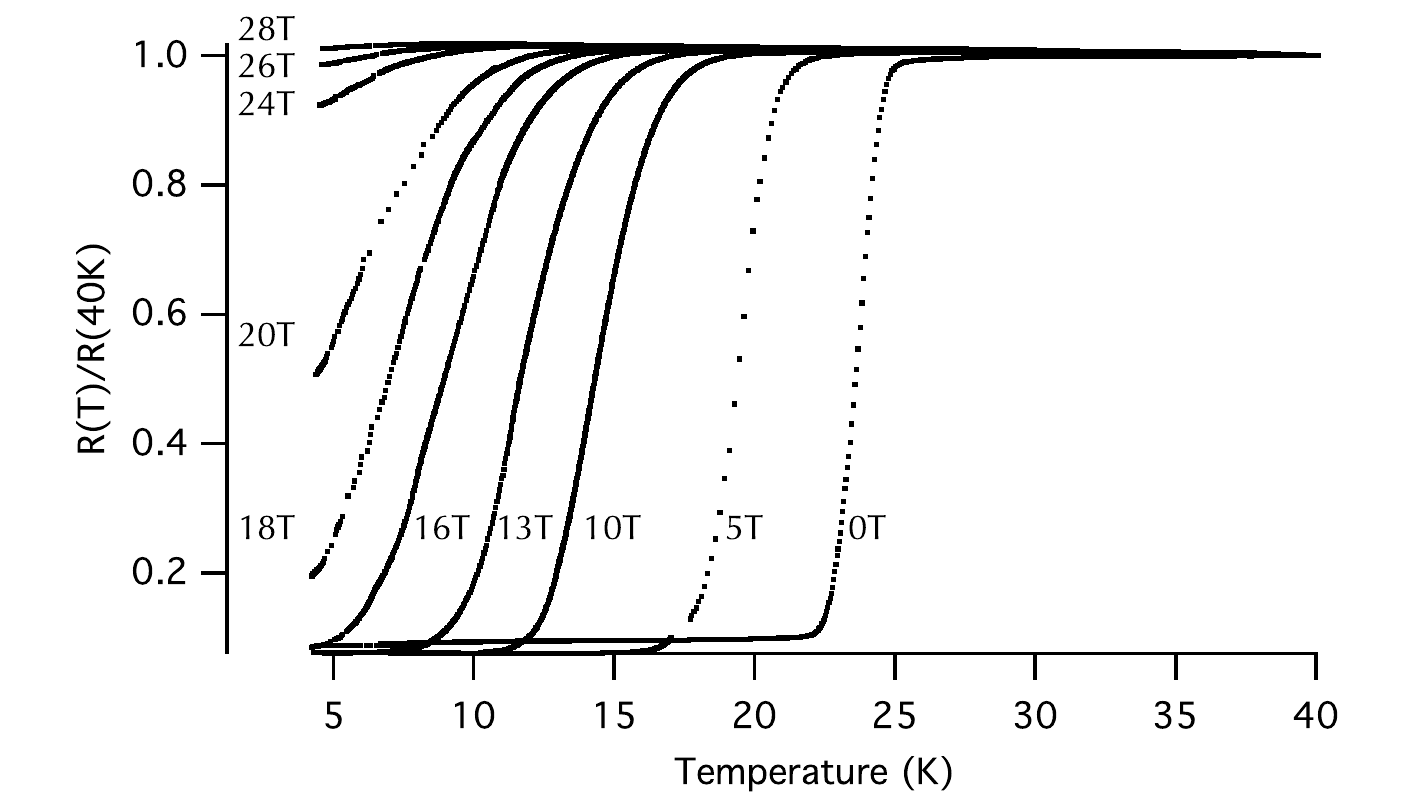}
  \caption{Transition curves of 60\,nm MgB$_2$ films with an applied magnetic field parallel to the film plane.}
 \label{fig:1}
\end{figure}
Because of the high vapor pressure of Mg and the low sputtering pressure of $2\times10^{-3}$\,mbar, the MgB$_{2}$ crystals exhibited a polycrystalline structure and a reduced critical temperature \cite{Liu:2001uga}. The critical temperature of our 60\,nm thick MgB$_{2}$ films was determined to be 24\,K with a specific resistance of $\rho\approx$ 150\,$\mu\Omega $cm at room temperature. The residual resistance ratio (RRR) was approximately constant at 1 for the entire temperature range down to the critical temperature. The resistance of the film was determinated to be essentially independent of temperature down to the critical temperature. This observation, together with the high resistivity  can be attributed  to a large number of scattering centers, thereby confirming the polycrystalline structure of the MgB$_{2}$ film. 
Although the sputtered MgB$_(2)$ are not epitaxial growth, the grains are not randomly oriented in thin sputtered MgB$_{2}$ surfaces. 

Prior investigations using $\Theta$-2$\Theta$ x-ray diffraction measurements have shown that sputtered MgB$_{2}$ crystals are perpendicularly oriented on the surface plane independent of the choice of substrate \cite{Fabretti:2012tx}. Using the Scherrer formula, a perpendicular grain size of 7\,nm could be estimated. However, despite the clear c-axis  MgB$_{2}$ (0002) peak, it was not clear whether most of the grains were regularly or randomly oriented. Therefore, it was necessary to conduct field-cooling experiments that were dependent on the orientation of the applied magnetic field.

The transport measurements were performed by using a $^{4}$He cryostat in a high-magnetic-field laboratory in Nijmegen (NL). The samples were cooled to 4.2\,K by an applied magnetic field perpendicular and parallel to the a-b substrate plane.
\begin{figure}
\includegraphics[width=8cm]{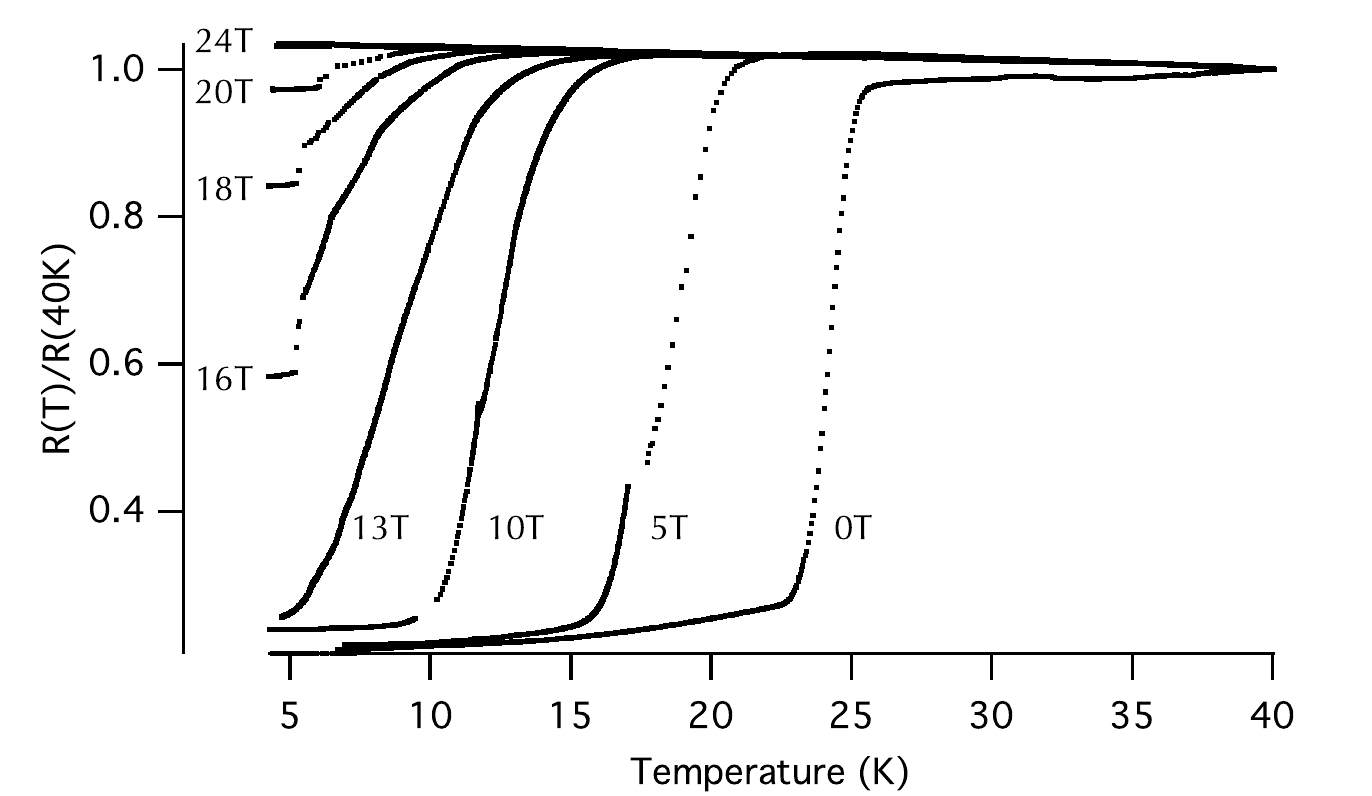}
  \caption{Transition curves of 60\,nm MgB$_2$ films with an applied magnetic field perpendicular  to the film plane.}
 \label{fig:2}
\end{figure}

Figure \ref{fig:1} shows the field-cooling transport measurements for a 60\,nm thick MgB2 sample with a magnetic field applied parallel to the film plane, and figure \ref{fig:2} shows the curves corresponding to a magnetic field applied perpendicular to the film plane. The critical temperature was defined by the 90\% criteria at the normalized resistance at 40\,K. For magnetic fields up to 16\,T for the perpendicular case and up to 18\,T for the parallel case, a residual resistivity was observed at 4.2\,K because of an increase of the transition width dependent on the applied magnetic field. However, because of the strong decrease in the resistance within a small temperature interval, we assume that the samples are approaching the phase transition.

\begin{figure}
\includegraphics[width=8cm]{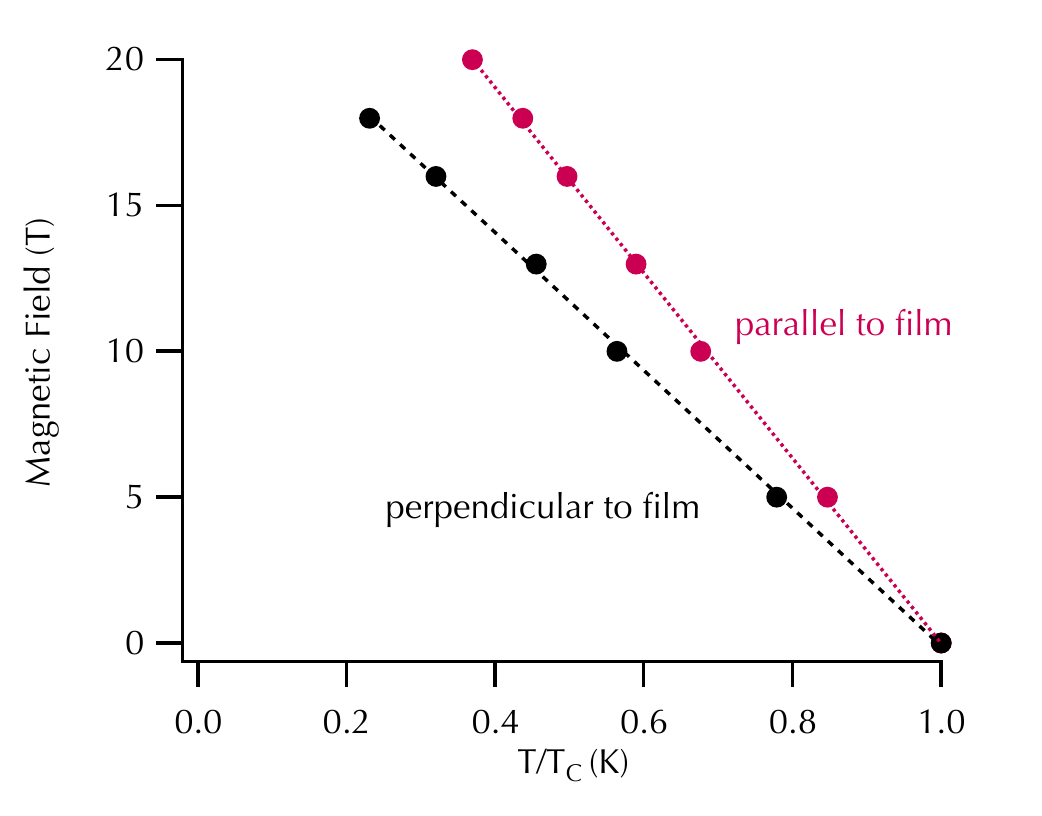}
  \caption{The estimated upper critical field at 0\,K. Using the 90\% criteria. The line is a linear fit to the data. Due to the constant slope the anisotropy ratios is still constant}
 \label{fig:4}
\end{figure}

Figure \ref{fig:4} shows the approximate dependence of the upper critical field H$_{c2}$ on the normalized transition temperature. These data points could be well described by a linear fit. Usually, the upper critical field can be estimated using the Ginzburg-Landau relation for a superconductor by assuming the dirty-limit extrapolation \cite{Gurevich:2003dd}. 
\begin{equation*}
H_{c2}=-0.69T_{c}\frac{dH_{c2}(T)}{dT}\mid_{T_{c}}
\end{equation*}

The upper critical field H$_{c2}$(0) for the a-b-axis direction is estimated to be 22.18\,T, and the H$_{c2}$(0) in the c-axis direction is  16.37\,T. These values imply an anisotropy factor $\gamma$ of 1.35 at 0\,K. 

Using Ginzburg-Landau theory and these critical fields, the coherence length at 0\,K can be estimated from a magnetic field applied parallel to the c axis with 
\begin{equation*}
H_{C2}^{c}=\frac{\phi_{0}}{2\pi\xi_{ab}^{2}}
\end{equation*}
and a magnetic field applied parallel to the a-b plane with 
\begin{equation*} 
H_{C2}^{ab}=\frac{\phi_{0}}{2\pi\xi_{ab}\xi_{c}} 
\end{equation*} 
where $\phi_{0}=2.07\cdot10^{-15}$\,Vs is the magnetic flux quantum. By using these formulas, the coherence lengths can be estimated to be $\xi_{ab}$=4.48\,nm and, for the perpendicular c axis, $\xi_{c}$=3.31\,nm. 

However, these values are underestimated, as has previously been shown by Gurevich for 400\,nm films \cite{Gurevich:2003dd}. Furthermore, the behavior of the curvature of the upper critical fields differs for single crystals in the clean limit and polycrystalline samples in the dirty limit. Lyard et al.\ have observed a positive curvature close to T$_{c}$, which becomes negative with a saturation at higher fields for an upper critical field applied perpendicular to the a-b plane of a single crystal with dimensions of 50x50x10\,$\mu$m$^{3}$ \cite{Lyard:2002id}. Theoretical calculations by Dahm et al. describe this behavior for single crystals \cite{Dahm:2003jl}. 

Due to the polycrystalline structure and therefore the many scattering centers of our film, this saturation could not be observed by our measurements. Therefore, we fit our measured points linearly and compared the results to a magnetic sweep measurement.

A linear extrapolation for both fields resulted in 31.74\,T for the parallel case and 23.5\,T for the perpendicular case at 0\,K. With these values, the anisotropy ratio was again found to be 1.35. Although there are different approaches to estimating the upper critical fields, the calculated anisotropy ratios are the same because of the linear fit. Despite the agreement of the anisotropy factor, the estimated upper critical fields are strongly different. However, as a consequence of the fact that both slopes of the fit curves are constant, the anisotropy factor of 1.35 is also constant over the entire critical temperature range. Only in the region of small applied fields should the anisotropy factor increase.

\begin{figure}
\includegraphics[width=8cm]{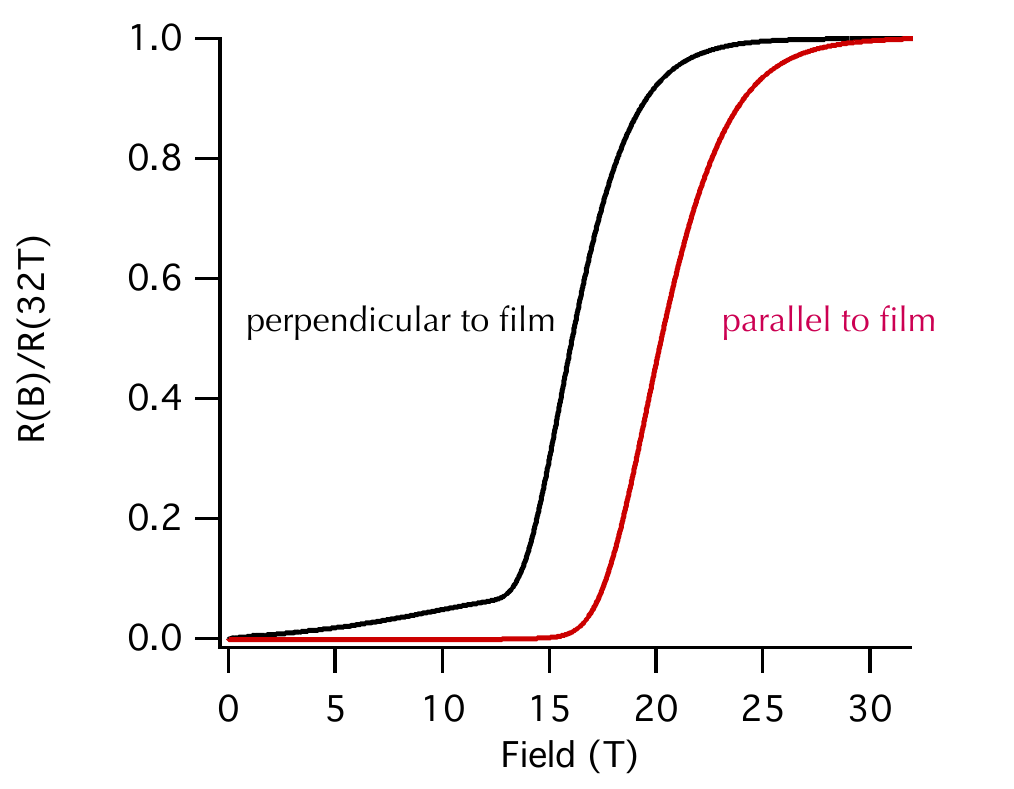}
  \caption{Sweep up to 32\,T at 4.2\,K. The slope up to 14\,T in the perpendicular case is caused by the different crystal sizes.}
 \label{fig:6}
\end{figure}
To confirm this value of the anisotropy ratio estimated above, measurements were performed for a sweep of the magnetic field at a constant temperature of 4.2\,K, as shown in figure \ref{fig:6}. For the linear case, the critical fields of the fitted slope should agree very well with the measured sweep. This type of measurement has the advantage that the phase transition, and therefore the anisotropy ratio $\gamma$, can be directly observed as a function of the applied magnetic field. The disadvantage is that the values are limited by the constant temperature of 4.2\,K. 

In the non-linear case, a strong deviation from the linear extrapolation should be observed. By using the 90\% criteria, the critical field in the perpendicular case was found to be 19.48\,T, which represents a deviation of less than 1\% from the estimated field of 19.6\,T at 4.2\,K. However, the deviation in the parallel case was found to be 26.28\,T, which represents a deviation of 3\,T compared to the linear approximation. Here, the linear extrapolation is clearly overestimated at lower temperatures, so  saturation could be expected in the parallel case for temperatures near 0\,K, as suggested by Gurevich. Therefore, a decrease in $\gamma$ is common for temperatures below 4.2\,K and samples in the dirty limit.

The curves of the sweeps exhibit a transition width of approximately 5\,T, which indicates an apparent retardation of the superconductivity on several grains as a function of the magnetic field. For clean single crystals, a small upper critical field of $\approx$4\,T for an applied field perpendicular to the film plane and $\approx$15\,T for an applied field parallel to the film plane is expected, which would result in an anisotropy ratio of up to 3.75 at 0\,K. Any impurity results in an increase of the upper critical fields in both directions, which has previously been observed by Gurevich et al.\ on polycrystalline MgB$_{2}$ samples \cite{Gurevich:2003dd}.

However, the constant distance of the sweep curves between the parallel and perpendicular magnetic fields with respect to the resistivity demonstrates that the grains are distributed homogeneously, i.e., the anisotropy ratio is nearly constant for the grains, independent of the transition temperature. 

Only at approximately 10\% of the resistance, the slope of the curve corresponds to the field perpendicular to the film plane becomes stronger than in the parallel case. This behavior could be caused by different crystal sizes, which was also observed by Lyard et al.\ \cite{Lyard:2002id}.

\begin{table}
\begin{tabular}{l|ccc}
film thickness (nm) & 30 & 60 & 120\\
anisotropy ratio $\gamma$ & 1.32 & 1.21 & 1.15\\
\end{tabular}
  \caption{The measured anisotropy ratio vs.\ film thickness using the 90\% criteria at 4.2\,K (The previously presented values were fits to 0\,K).}
 \label{tab:1}
\end{table}
Table \ref{tab:1} shows how the anisotropy factor at 4.2\,K depends on the film thickness. A decrease in the anisotropy factor of up to 12\% was observed. The decrease of the anisotropy factor can be explained by an increase of the crystal size. Handstein et al.\ observed an anisotropy ratio of 1.1 for imperfect hot deformed bulk samples, and an anisotropy factor of 1.8 to 2 was observed by Patnaik et al.\ for c-axis-oriented thin films \cite{Patnaik:2001vd, Handstein:2001ts}. However, an anisotropy factor of up to 9 was observed by Simon et al.\ for randomly oriented powder samples \cite{Simon:2001gh}. Our values are comparable to the values of Jung, who predicted an anisotropy factor of 1.25 for thin films \cite{Jung:2001wf}.
%
%

In summary, we fabricated sputtered polycrystalline MgB$_{2}$ films in which most of the grains are c-axis orientated. The anisotropy ratio was nearly constant in the investigated temperature and field range, which was determined using a linear extrapolation as well as the Ginzburg Landau theory (at 0\,K). Overall, the thin films in the thickness regime between 30\,nm and 120\,nm are an ideal starting point for further experiments utilizing Meservey-Tedrow tunneling. Textured films enable us to tunnel preferably in the $\pi$ as well as the $\sigma$-band of the two-band superconductor MgB$_2$.

%

We would like to acknowledge the MIWF of the NRW state government and the German Research Foundation DFG for financial support and we are very grateful to J.S. Moodera and G.\ Reiss for  encouraging us to start this project. We acknowledge support of the HFML-RU/FOM, member of the European
Magnetic Field Laboratory  (EMFL). Part of this work has been
supported by EuroMagNET II under  EU contract number 228043.%
\bibliography{LibBib}
\end{document}